\documentclass
[prl, aps, onecolumn,showpacs]{revtex4}

\usepackage{psfrag}
\usepackage{graphicx}
\usepackage{graphics}
\usepackage{bm}
\usepackage{color}
\usepackage{centernot}

\newcommand{\be}{\begin{equation}}
\newcommand{\ee}{\end{equation}}
\newcommand{\ba}{\begin{eqnarray}}
\newcommand{\ea}{\end{eqnarray}}

\newcommand{\dcom}[1]{}
\newcommand{\dnote}[1]{}

\newcommand{\gsim}{\raise.3ex\hbox{$>$\kern-.75em\lower1ex\hbox{$\sim$}}}
\newcommand{\lsim}{\raise.3ex\hbox{$<$\kern-.75em\lower1ex\hbox{$\sim$}}}

\begin{document}

\renewcommand{\thefootnote}{\fnsymbol{footnote}}


\renewcommand{\thefootnote}{\arabic{footnote}}
\setcounter{footnote}{0} \typeout{--- Main Text Start ---}

\title{ Variable Modified Newtonian Mechanics III: Milky Way Rotational curve}
\author{ James C.~ C. Wong}
\affiliation{Department of Electrical and Electronic Engineering, University
of Hong Kong. H.K.}

\date{\today}
\begin{abstract}
For a point mass residing in an expanding universe, within General Relativity (GR), a new metric \cite{wong} is found to lead to a cosmological background dependent MOND-like acceleration in addition to the Newtonian acceleration. In \cite{wong2}, we study the monolithic evolution of an spherical overdensity at recombination in this combined acceleration, called VMOND. Under reasonable relaxation assumptions we find that a massive spherical galaxy  with a stable core can form at $z\geq 7$.  For galaxy mass $M=10^{10.5}M_{\odot}$ and a realistic initial overdensity, the model late time MOND acceleration $a_0^{VM}(r)$ at radius $r$ takes on values similar to the canonical MOND acceleration $a_0$ at large radius $r$.
\\\\
In this work, we consider an idealised model of rotating galaxy formation in which a Milky Way mass overdensity under VMOND monolithically evolves into a virialised sphere.  We assume that this virialised sphere is given an uniform systematic angular velocity, which equilibriates to a flat disk according to Mestel's analysis \cite{mestel}.   We apply the Mestel's disk potential due to the flatten virialised sphere under VMOND to calculate the rotational curve at radius $17.77kpc-27.30kpc$. We find that the model combined acceleration leads to a Baryonic Tully-Fisher Relation (BTFR) with radius dependent acceleration $a_0^{VM}(25kpc)\sim O(a_0)$. The model rotational velocity in the same radius range matches Gaia DR3 measurements \cite{eilers}-\cite{ou} very closely.
\end{abstract}

\pacs{??}

\maketitle
\section{Introduction}
\noindent Gas rich galaxies have non-Newtonian rotational velocities whose asymptotic behaviour can be parametrisesd approximately \cite{mcgaugh2004,lelli,milgrom0,sanders0} by the empirical Baryonic Tully-Fisher Relation (BTFR)
\begin{equation}
V = (a_0G M_{Gal})^{1/4},
\label{BTFR}
\end{equation}
which falls into the general class of well known dynamical properties of galaxies called Radial Acceleration Relations (RAR) \cite{lelli}.
\noindent A common response is to retain General Relativity (GR) and therefore Newtonian behaviour by postulating the existence of electromagnetically invisible cold Dark Matter (DM) to bridge the gap. The $\Lambda CDM$ model is found most successful at explaining the gravitational potential from radiation-matter equality \cite{mukhanov}-\cite{planck} to reproduce the cosmic microwave backgorund (CMB) angular power spectrum. Apart from the long standing difficulties \cite{vand,muller,kroupa2014,bowman,mcgaugh2018}, serious challenges to the $\Lambda CDM$ model emerges recently. At large scales these include the Hubble Tension \cite{riess}-\cite{valentino}, the $\sigma_8$ Tension  \cite{bohringer}-\cite{einasto} and Phantom dynamical dark energy from DESI DR2 data \cite{karim}-\cite{scherer}. At galactic scales, the CDM model does not anticipate the early emergence of massive galaxies \cite{mcgaugh2024} and early time Supermassive Black Holes \cite{cho}, both  require a period of significant local gravitational potential  increase for overdensity growth after recombination or super Eddington accretion rate of an early time black hole seed \cite{smith}.
\\\\
Another approach is to construct variants of Newtonian gravity to change long-range behaviour to avoid the need for additional particles.
The most familiar example of this approach is Modified Newtonian Dynamics (MOND) in which Milgrom \cite{milgrom} modifies the
gravitational acceleration to $\sqrt{g_Na_0}$, when the Newtonian acceleration $g_N$ reaches values far below a phenomenologically acceleration scale $a_0$. Although the BTFR in Eq.(\ref{BTFR}) does not require $a_0$ to be a constant of nature a priori, before an underlying mechanism which produces $a_0$ is identified, assuming a constant (canonical) scale $a_0=1.2\times 10^{-10} ms^{-2}$ using the observed value (with small scatter) of Gas rich galaxies \cite{milgrom,famaey2013,mcgaugh} becomes a successful paradigm for studies at galactic scales without the need for dark particles. However, outside of galactic scales $10^7M_{\odot}-10^{11}M_{\odot}$, canonicity of $a_0$ becomes less convincing. At late time solar system scale \cite{tremaine}-\cite{desmond} and Wide-Binary scale \cite{banik}, canonical $a_0$ has met with observation difficulties. Non canonical acceleration value ($7.2a_0-20a_0$) is also preferred at large graviationally bound systems such as Bright Cluster Galaxy (BCG) \cite{tian}. In galaxy clusters, a MOND acceleration treatment of the high temperature X-ray region (under $300kpc$)  \cite{aguirre,sanders} shows that either more
invisible mass is required or a fourfold {\it increase} of $a_0$ to produce the observed temperature is necessary. Further work shows that \cite{mcgaugh2020}-\cite{li2023} at larger radius observations  favour a MOND acceleration  at $10a_0-20a_0$ before the corresponding MOND acceleration falls below the canonical $a_0$ value at very large distances ($2-3Mpc$). Introducing sufficient invisible mass at the core to lift the canonical MOND velocity dispersions to match the small distance observation leads to MOND velocity dispersion predictions far larger than observations at 1$Mpc$ scale \cite{lelli2024}. Even then, large scale structure formation simulations need a significantly {\it smaller} $a_0$ than the canonical $a_0$ value \cite{nusser2002}. At smaller scales, the Tidal Dwarf which is formed at late time is dynamically Newtonian \cite{lelli2}. High quality Milky Way rotational curve data, \cite{ou} also challenges the canonical MOND paradigm\cite{chan}-\cite{coquery}. As a result, the MOND phenomenology (Tully-fisher relation and Faber-Jackson relation) indicates that the MOND acceleration $a_0$ is not a canonical value. To paraphrase the words of \cite{monjo1}, the canonical MOND paradigm highlights the strong coupling between total gravitational acceleration with the baryonic central mass, but is incomplete in its lack of flexibility to account for the wide variety of observed phenoma. Due to the resemblance of $a_0$ value with the cosmological acceleration $\sim cH_0$, Milgrom has considered a cosmologically varying $a_0(z)$ \cite{milgrom2015}. 
\\\\
In any case, the canonical MOND
remains a limited phenomenological model restricted by its non-relativistic formulation \cite{mcgaugh,famaey} and as it stands does not provide the non-baryonic gravitational potential to enhance the CMB acoustic peaks \cite{mcgaugh} and to dampen baryonic acoustic oscillations after recombination.
The above difficulties and others
have led to a proliferation of variants of the original model; in particular, Bekenstein TeVeS \cite{bekenstein}, EMOND \cite{zhao}, GMOND \cite{khoury2015}, Emergent
Gravity \cite{verlinde} and MOG \cite{moffat2}-\cite{moffat} and Relativistic MOND \cite{skordis}-\cite{skordis1}, Hyperconical model \cite{monjo2} all involving a non-Newtonian acceleration obtained by modifying GR. Apart from MOG, Relativistic MOND and Hyperconical model, all of these models are ruled out by Gravitational wave speed observation \cite{boran}. In MOG \cite{moffat1}) and Hyperconical model \cite{monjo1}, modification of GR leads to a cosmic background dependent acceleration $a_{\gamma, 0}=2\gamma_0^{-1} c t^{-1}$, where $\gamma_0\sim 12-13$ which is either obtained empirically or derived from the model assumption and $t$ is the age of the universe, which depends on the underlying cosmology with or without cold dark matter. If we rewrite $a_{\gamma,0} =cH_0 \gamma_0^{-1}(z)$, where $\gamma_0(z)$ is a redshift dependent number, together with Milgrom's $a_0(z)$ \cite{milgrom2015}, these models   favours a variable MOND that would take the form $cH_0 \gamma_0^{-1}(z)$.
\\\\
The challenge of a successful MOND variant is that broadly speaking it needs to be a relativistic theory which provides a non-baryonic gravitational potential from radiation-matter equality to the beginning of matter dominance to produce the observed CMB angular power spectrum and the suppressed Baryonic acoustic oscillations. It should also produce a variable MOND acceleration $a_0$ (or better a MOND-like potential) which explains the various mass discrepancies up to 1 Gpc \cite{kroupa}.
\\\\
In the work of \cite{wong}-\cite{wong2}, without modifying GR we find a new solution of the Einstein equation for a point mass in an expanding background, which gives rise to a MOND-like acceleration $\ddot{r}=\sqrt{g_N (\frac{1}{2}H^2r)}$ for a background Hubble parameter $H$ and radius $r$. (Here $\frac{1}{2}H^2r\simeq \frac{1}{2}H\dot{r}$ for $r$ large, which can take the form $cH_0 \gamma_0^{-1}(z,r)$.) At late time $H=H_0$ and Oort cloud scale $r<10^5 AU$, this acceleration is far weaker than Newtonian acceleration and avoids the problem confronting canonical MOND. In the Appendix below, we include calculations at Solar system and Wide Binary scale, which show that VMOND can avoid the difficulties encountered by canonical MOND at small scales.
\\\\
 For a spherical overdensity at recombination, the MOND-like acceleration leads to a much faster overdensity growth and turns around at very high redshift \cite{wong}-\cite{wong2}. Here the MOND-like acceleration $\frac{1}{2} H^2r$ is far stronger than the same MOND-like acceleration in the bounded solar system which is formed at late time. After turnaround, individual mass shell picks up appropriate angular momentum which is either systematic or non-systematic.  In a collisionless violent relaxation, the overdensity with non-systematic angular  momentum gravitationally evolves to a virialised sphere without significant potential energy loss so that the non-Newtonian potential retains its strength (or depth) at the turnaround redshift $z_{ta}$. The Virial theorem stipulates that near equilibrium, the virialised potential is dominant and the average kinetic energy approaches its equilibrium values. The modelled averaged Velocity dispersion-Radius relation is found compatible with observations in \cite{durazo}. For elliptical galaxy with Milky Way mass $10^{10.5} M_{\odot}$ and BCG mass $10^{11.5} M_{\odot}$, the corresponding $a_0^{VM}$ values at large radius are found to match observations. 
\\\\
The modelling of a nearly uniform Newtonian sphere with uniform systematic angular velocity (angular momentum is conserved at its particle level ) which flattens into a disk has been considered analytically by Mestel \cite{mestel}, who obtains an equilibrium disk potential which differs from that of Newtonian potential of a sphere.
\\\\
In this work, we consider a simple model of rotating galaxy formation, in which a Milky Way mass uniform overdensity after recombination evolves monolithically into a virialised sphere under VMOND as considered in \cite{wong2}. Following Mestel, we assume that an uniform systematic angular velocity is given to the virialised sphere such that this sphere equilibriates into a rotating thin disk with a well defined potential \cite{mestel}. We then calculate the rotational curve in the radius range where non-Newtonian effect becomes dominant and compare to recent measurements \cite{eilers}-\cite{ou}.
\\\\
In section 2, for completeness we recapitulate the key features that this new metric entails, the overdensity growth rate under VMOND acceleration and the corresponding virialised spherical potential.  In section 3, we borrow the results from \cite{mestel} to obtain the virialised disk potential and calculate the VMOND acceleration $a_0^{VM}$ for the virialised disk. We also calculate the rotational velocity at radius $17.77kpc-27.30kpc$ and compare with recent measurements \cite{eilers}-\cite{ou}. Section 4 is a summary and discussion. In the Appendix, we calculate the particle orbit formed at the present epoch under the VMOND potential and also calculate the VMOND effect at the Wide-Binary scale.
\section{2.1: The Model}
\noindent In \cite{wong}, we notice that in the Tolman-Lema$\hat{i}$tre spherical symmetric metric formulation, specifying the free falling velocity of a particle specifies the metric. The particle free falling velocity around a central point mass is, according to Newton's law,
\begin{equation}
\dot{r} =-\sqrt{\frac{2GM}{r} }
\end{equation}
where $G$, $M$ and $r$ are the Newton's constant, the central point mass and the radial
distance respectively. From this free falling speed one obtains the Schwarzschild  Lema$\hat{i}$tre metric with coordinate time $\tau$, comoving distance $\varrho$ ($d\Omega^2=d\theta^2+sin^2\theta d\varphi^2$)
\begin{equation}
ds^2=c^2d\tau^2-\frac{2GM}{c^2r}d\varrho^2-r^2d\Omega^2.
\end{equation}
This metric can be transformed to the more familiar {\it Schwarzschild metric} in curvature coordinates.
\\\\
If we specify the free falling speed of a particle in a cosmlogical background with Hubble parameter $H(z)$ at redshift $z$, according to Hubble's Law,
\begin{equation}
\dot{r}=H(z) r,\:\:\: H(z) =\frac{1}{a(\tau)} \frac{d a(\tau)}{d\tau},
\end{equation}
where $a(\tau)$ is the scale factor at $z$ and we obtain the Friedmann-Lema$\hat{i}$tre metric
\begin{equation}
ds^2=c^2d\tau^2-a^2 d\varrho^2-r^2d\Omega^2.
\label{frw}
\end{equation}
For a central point mass in an expanding background, in \cite{wong} we find a new metric having the form 
\begin{equation}
ds^2=c^2d\tau^2-\frac{2GMa^3}{c^2r}d\varrho^2-r^2d\Omega^2.
\label{LT1}
\end{equation}
which, for a weak gravitational potential, is equivalent to the perturbed FRW metric in Conformal Newtonian gauge. Here the underlying free fall velocity is given by
\be
\dot{r}= H(z)r-\sqrt{\frac{2GM}{r}},
\label{dotr1}
\ee
Eq.(\ref{dotr1}) depicts that the particle will follow the Hubble law at large distances, but at small distances it will follow a Newtonian free falling velocity. This equation differs from the $\dot{r}$ choice in Schwarzschild de-Sitter metric,
\begin{equation}
\dot{r}^2=\frac{2GM}{r}+\frac{\Lambda c^2 }{3}r^2,
\end{equation}
where $\Lambda$ is the cosmological constant.  At slow speeds in radial direction, Eq.(\ref{dotr1}) is formally the same as the equation of motion in Newtonian perturbation theory
\begin{equation}
\dot{r}=H(z)r+v_p,\:\:\:\:\ddot{r} =\frac{\ddot{a}}{a}r+\dot{v}_p,
\label{NPT}
\end{equation}
where $v_p$ is the peculiar velocity of the particle. Eq.(\ref{NPT}) is the base equation used to obtain the overdensity evolution in the Newtonian perturbation theory \cite{mukhanov}.
Contrary to the Newtonian perturbation theory due to the radius dependence in $v_p=-\sqrt{2GM/r}$, Eq.(\ref{dotr1}) in fact leads to the acceleration equation 
\begin{equation}
\frac{d \dot{r}}{dt}=\dot{r} \bigg(\sqrt{\frac{GM}{2r^3}}+H\bigg)+r\dot{H},
\label{ddotr0}
\end{equation}
\begin{equation}
\ddot{r}=-\frac{GM}{r^2} -\sqrt{\frac{H^2r}{2}}\sqrt{\frac{GM}{r^2}}+\frac{\ddot{a}}{a}r =-\sqrt{\frac{GM}{r^2}a_0^{VM}} +\frac{\ddot{a}} {a}r,.
\label{nNewton1}
\end{equation}
We term the sum of the Newtonian and the non-Newtonian gravitational acceleration  the "VMOND" acceleration, where $a_0^{VM}$ corresponds to the $a_0$ in the canonical MOND paradigm at large distance limit.
The radial acceleration including angular momentum is given as
\begin{equation}
\ddot{r}=  \frac{h^2}{r^3} -\frac{GM}{r^2}-\sqrt{\frac{1}{2}H^2(r)r}\sqrt{\frac{GM} {r^2} } +\frac{\ddot{a}} {a}r,
\label{vmond}
\end{equation}
where $h$ is the angular momentum per unit mass.  
The slow speed energy equation is given by integrating Eq.(\ref{vmond})
by
\begin{equation}
\frac{1}{2}\dot{r}^2+\frac{h^2}{2r^2} =E+\frac{1}{2}\bigg(\sqrt{\frac{2GM}{r}}-H(z)r\bigg)^2=E+\alpha\frac{GM}{r},\:\:\:\alpha =\bigg(\sqrt{\frac{\rho_H(z)}{\rho_b(r)}}-1\bigg)^2,
\label{potl44}
\end{equation}
where $E$ is the energy of the particle. 
The distance from the centre
\be
r_{ta}=\bigg(\frac{2GM}{H^2}\bigg)^{1/3},
\label{rh}
\ee
which describes a turnaround radius of the particle. At late time solar system central mass, $r_{ta} \sim 10^{7}AU$, but the solar system scale is $10^{5} AU$. This means that within solar system, the dominant gravitational acceleration remains Newtonian.
\subsection{ 2.2: Density perturbation evolution}
The evolution of an overdensity from recombination is presented in the first paper of this series \cite{wong}, here we recall some of the main features which will be relevant for our study.
\\\\
The Hubble parameter at redshift $z$ where radiation is negligible is given by the Friedmann Equation,
\be
H^2(z) =\frac{8 \pi G}{3}\rho_H=\frac{8\pi G}{3} \bigg(\rho_b+\rho_{\Lambda}\bigg)=H_0^2 \bigg(\Omega_b(1+z)^3+\Omega_{\Lambda}\bigg),\:\:\:H^2_0=\frac{8 \pi G}{3}\rho_c,
\label{H22}
\ee
where $\rho_b$ and $\rho_{\Lambda}$ are cosmological background densities of baryonic matter and dark energy respectively. $H_0$  and $\rho_c$ are the Hubble parameter and the critical density of the present epoch. $\Omega_m$ and $\Omega_{\Lambda}$ are the density parameter for matter and $\Lambda$ respectively. \\
Note: Although there is another way to define the Hubble parameter which can explain the dark matter potential at early times \cite{wong3}, this alternative prescription has a matter dominant epoch at $z=600-1.67$ and thus performs in this epoch effectively in the same way as the Friedmann-Lema$\hat{i}$tre-Robertson-Walker (FLRW) cosmology Eq.(\ref{H22}).
\\\\
Soon after recombination, the cosmological background is in the matter dominant epoch with mean density 
\be
\rho_b=\frac{1}{6\pi Gt^2}.
\ee
The baryon density perturbation $\delta$ is specified by 
\begin{equation}
\delta = \frac{ \rho-\rho_b} {\rho_b}=\frac{\delta\rho_b}{\rho_b},
\label{delta}
\end{equation}
where $\rho$ is the total mass density of the overdense region inside radius $r$. The baryon matter overdensity is given by $ \delta \rho_b$.
\\\\
In the matter dominant era, the Einstein equation (Poission Equation) for Eq.(\ref{LT1}) involving mass density is 
\be
\nabla^2 \Phi =4\pi G\rho=4\pi G (\delta \rho_b+\sqrt{\delta \rho_b \rho_b} +\rho_b)=4\pi G (\delta+\delta^{1/2}+1)\rho_b.
\ee
We note that the total density is therefore $\rho=(1+\delta+\delta^{1/2})\rho_b$, where total perturbation is $\Delta(\delta) =\delta+\delta^{1/2}$. The "dynamical" mass density $\delta^{1/2}\rho_b$ corresponds to the non-Newtonian acceleration in the peculiar acceleration. At large redshifts when $\delta\ll1$, the dynamical mass density, which is invisible electromagnetically, dominates over the baryonic overdensity. 
\\\\
Following the Newtonian perturbation approach  \cite{binney} based on collisionless Boltzmann equation, the Euler equation and continuity equation requires differentiation w.r.t. to time before combining with the Poisson equation to give the (second order in time derivative) overdensity evolution equation. In the Euler equation, differentiating the peculiar velocity in the cosmic background leads also to the 
additional non-Newtonian acceleration which corresponds to dynamical energy density . Therefore, $\Delta(\delta) =\delta+\delta^{1/2}$ becomes the  effective overdensity that goes into the evolution equation 
\begin{equation}
\frac{\partial^2}{\partial t^2}\Delta(\delta) +2H\frac{\partial}{\partial t} \Delta(\delta) = 4\pi G \rho_b \Delta(\delta).
\label{deltadot22}
\end{equation}
In a flat universe at the matter dominant epoch where $H=2/(3t)$, Eq. (\ref{deltadot22}) has the growth solution
\begin{equation}
\Delta(\delta)=\delta +\sqrt{\delta} \propto  t^{2/3} \propto a(t);\:\:a(t)\propto \frac{1}{1+z}.
\label{Delta1}
\end{equation}
In the VMOND paradigm, at $\delta\ll1$, we obtain a new $\delta$ growth rate $\delta \propto t^{4/3} \propto a(t)^2$. In the MOND paradigm, the same overdensity growth rate $\delta\propto a^2$ is used in large scale structure simulation \cite {nusser2002} and in elliptical galaxy formation simulation \cite{sanders2007}. 
\\\\
From an initial overdensity $\delta_{int}$ at recombination, we can use $\delta+\delta^{1/2}\propto a(t)$ to calculate the overdensity $\delta$ at redshift $z<1080$ by
\begin{equation}
\delta+\sqrt{\delta}=\bigg(\delta_{int}+\sqrt{\delta_{int}}\bigg) \bigg(\frac{1081} {1+z}\bigg)=\frac{A_0}{1+z},\:\:\: A_0 =\bigg(\delta_{int}+\sqrt{\delta_{int}}\bigg)1081.
\label{dpsd}
\end{equation}
where $A_0$ is fixed by the value of $\delta_{int}$. In a cubic box approximation, at characteristic box size scale $R$ the overdemsity behavies as $\delta (R) \propto R^{-S},\:\:0<S<3$ \cite{binney} (pp720-721),\cite{nusser2005}. In simple overdensity evolution model, one would use a larger initial uniform overdensity value to produce small galaxy \cite{bromm1} and smaller initial uniform overdensity to produce larger galaxy \cite{bromm1},\cite{bromm2}. 
\\\\
The baseline CMB average temperature variation $\frac{\delta T}{T}=1\times 10^{-5}$ corresponds to an initial baryon overdensity $\delta_{int} =3 \times 10^{-5}$ and $A_0=5.94$. Large galaxies will require a larger initial overdensity. In \cite{sanders2007}, to simulate the evolution of a galaxy with mass $10^{11}M_{\odot}$ Sanders takes $\delta_{int} =1.8\times 10^{-3}$ which corresponds to $A_0=47.8$. In a direct collapse model of a DM with baryon halo with $M=10^8M_{\odot}$, an initial overdensity $\delta_{int} \sim 10^{-3}$ is used \cite{bromm2} .
\\\\
In \cite{wong2}, we choose a slightly higher value $\delta_{int}=2.8\times 10^{-3}$ ($A_0=60$) to develop into a $10^{10.5}M_{\odot}$ elliptical galaxy. We obtain the turnaround redshift $z_{ta}$ where $\delta=1$ from Eq.(\ref{dpsd}),
\begin{equation}
1+z_{ta}=\frac{A_0}{2}=30, \:\:\: z_{ta}=29.
\label{ta}
\end{equation}
%
if we work with the Newtonian gravity only evolution where $\Delta=\delta$, we obtain
\begin{equation}
1+z_{ta}= 1081\delta_{int} =3.02, \:\:\:z_{ta}=2.02.
\end{equation}
We can see the VMOND potential significantly lifts $z_{ta}$ to a much higher redshift, which could provide the model basis to address the issue of massive galaxies at high redshifts. 
\\\\
 From Eq.(\ref{ta}), we can estimate the turnaround time when the overdensity reaches $\delta=1$.
In a matter only universe, we have the scale factor $a(t) \propto t^{2/3}$, so that
\begin{equation}
H(z)=\frac{\dot{a}}{a} =\frac{2}{3t}.
\label{mt1}
\end{equation}
At redshift $z_{ta}$, we can calculate the turnaound time $t_{ta}$   by
\begin{equation}
t_{ta}=\frac{2}{3H(z_{ta})}.
\end{equation}
After the turnaround, in \cite{wong2} we find the particle free fall time $t_{ff} =t_{ta}$ which is also the violent relaxation time scale \cite{lyndenbell}. In a violent relaxation scenario,  the time to different stage of virialisation $t_{vr}$ is given by \cite{wong2}
\begin{equation}
t_{vr}=t_{ta}+t_{ff}+ Nt_{dyn} =\frac{2}{3H(z_{ta})} \bigg(2+3\sqrt{2}N\bigg)=\frac{2}{3H(z_{vr})},
\label{tvr1}
\end{equation}
where $z_{vr}$ is the virialisation redshift and $N$ is the number of cycle time $t_{dyn}=2 \sqrt{\frac{r^3}{GM}}$ after $t_{ff}$ taken to attain an effective virialisation. From Eq.(\ref{tvr1}), we obtain
\begin{equation}
1+z_{vr}=\frac{1+z_{ta}}{(2+3\sqrt{2}N)^{2/3}}
\label{zvrzta}
\end{equation}
In the dark matter halo collapse model \cite{bromm2}, the virialised time is taken at $N=0$.
\\\\
We assume that after $t_{ff}$ ($N=0$), a tight (but messy) central configuration could occur. Eq.(\ref{zvrzta}) shows that this occurs at $z_{vr}=17.9$.
\\\\
For a Quasi-Stationary-State (QSS) obtained in simulation \cite{binney}, \cite{ciotti}, on taking $N=1$, QSS occurs at
\begin{equation}
z_{vr}=\frac{1+z_{ta}}{3.39}-1=7.85.
\end{equation}
An effectively complete virialisation occurs at $N=3$ \cite{ciotti}, where
\begin{equation}
z_{vr}=\frac{1+z_{ta}}{6.0}-1=3.88.
\end{equation} 
The analysis above suggests that given the chosen $\delta_{int}$, the VMOND potential is sufficient to provide for a tight minimum galactic configuration at $z_{vr}\sim 17.9$ and a stable (Quasi-Stationary State) galactic core by $z\geq 7$. The central core will remain stable while evolves through the range $1.5<z<6.5$, as observed  in \cite{ferreira2}. 
 %
%
\subsection{2.3: Virialisation at high redshift and $a_0^{VM}$}
After an overdense cloud turnarounds and collapses gravitationally at high redshift to reach a central core, we expect that the cloud primarily relaxes through a "violent relaxation" similar to what is found in the three-dimensional dissipationless collapse simulation in a MOND potential by Nipoti et al. \cite{nipoti} and reaches a meta-equilibrium first and eventually fully virialises. 
\\\\
We start with the central mass upto a shell at $r$, with a power-law density $\rho(r)=\bar{\rho}r^{-S}$ ($\bar{\rho}$ is constant and $S>0$), which is
\be
M(R)=4 \pi \int_0^{R} r^2 \rho(r) dr =4 \pi \bar{\rho}\int_0^{R} r^{-S} r^2dr=\frac{4 \pi \bar{\rho}} {(3-S)} R^{(3-S)}.
\label{mt}
\ee 
Given the time averaged Kinetic energy $K=\frac{1}{2}M V^2$ (here $V$ is the average rotational speed) and average potential energy $\Psi$, the virialised relation is given by $2K+\Psi=E$, with the average energy $E\rightarrow 0$ at equilibrium.
\\\\
The virial potential energy is given by
\be
\Psi=-4\pi\int_0^{R} drr^3 \rho(r) \nabla \cdot \Phi,
\ee
where $\Phi$ is the local potential. The virial potential consists of both a Newtonian and a non-Newtonian potential
\be
\Psi=-4\pi \int_0^R dr r^3 \bar{\rho}r^{-S} \bigg(\frac{GM(r)}{r^2}+  H(z_{ta})\sqrt{ \frac{GM(r)}{2r}}\bigg)=-C_1(S) \frac{GM^2}{R}-C_2(S)MH(z_{ta})\sqrt{\frac{GMR}{2}}.
\label{psi}
\ee
\begin{equation}
C_1(S)=\bigg(\frac{3-S}{5-2S}\bigg) ;\:\:\:C_2(s)= \bigg(\frac{3-S}{5-\frac{3}{2}S}\bigg).
\end{equation}
The MOND-like virial potential comes through a collisionless relaxation and virialisation process, we can assume that there is no potential energy loss and keep the redshift "$z_{ta}$" in the evaluation of the (redshift dependent) virialised potential. This primeval turnaround redshift $z_{ta}$ value is crucial in providing the strong non-Newtonian gravity at galactic scales or above as we shall find below. We notice that the virialised potential differs from the point mass potential by the factors $C_1(S)$, $C_2(S)$ and the estimated turnaround redshift $z_{ta}$. It is worth noting that if the non-Newtonian potential is treated as a dark halo (potential), it can lead to an apparent dark halo-visible matter coupling relation \cite{salucci}, but this potential is not easily distinguishable from collisionless dark particle halo based on gravitational effects alone.
\\\\
For a stationary orbit where the kinetic energy is dominated by an averaged rotational velocity $V^2$, we have 
\be
V^2(R)=C_1(S)\frac{GM}{R}\bigg(1+\frac{C_2(S)} {C_1 (S) }\sqrt{\frac{H^2(z_{ta})R^3}{2GM}}\bigg) =C_1(S)\frac{GM}{R}\bigg(1+ \frac{C_2(S)} {C_1 (S) }\sqrt{\frac{\rho_H(z_{ta})}{\rho_b(R)}}\bigg),
\label{virialeq1}
\ee
where $\rho_H(z_{ta})$ is the cosmic background density at $z_{ta}$. At short distnce $R$, the baryon density $\rho_b(R)$ is high such that $\rho_b(R)\gg \rho_H(z_{ta})$, we have the nearly Newtonian behaviour from Eq.(\ref{virialeq1}). Baryonic mass conservation leads to 
\be
M=\frac{4\pi}{3} \rho_b(R)R^3=\frac{4\pi}{3}\rho(z_{ta}) r_{ta}^3.
\ee
This equation leads to
\be
\frac{\rho_b(R)}{\rho(z_{ta})}=\frac{\rho (z_{ta})}  {\rho (z_{ta} ) } n^3 =n^3,\:\:\:\: n=\frac{ r_{ta}}{R}.
\ee
which relates the rotational speed at $R$ with the turnaround radius $r_{ta}$.
We can rewrite Eq.(\ref{virialeq1}) as
\be
V^4(R)= C_2(S)^2GM \bigg(\frac{1}{2}H(z_{ta})^2R\bigg) \bigg(1+\frac{C_1(S)}{C_2(S)}\sqrt{\frac{\rho_b(R)}{\rho_H(z_{ta})}}\bigg)^2 =GMa_0^{VM}(z_{ta}, R).
\label{vsigma11}
\ee
At large $R$, $\rho_b(R)/\rho_H(z_{ta})$ approaches a number close to unity.
Eq.(\ref{vsigma11}) with an emphasis on the far R.H.S. can be recognised as the Faber-Jackson relation or Tully-Fisher relation with a variable $a_0^{VM}(z_{ta}, R)$. In \cite{wong2}, we find that the virialised potential of a sphere has a Newtonian  dominant central region around its half mass radius and a MOND like region at large radus, this is compatible with the findings in \cite{durazo}. We also find that a Milky Way mass ($10^{10.5}M_{\odot}$) overdensity $\delta=2.8\times 10^{-3}$ at recombination will result in a corresponding VMOND acceleration which takes the value $a_0^{VM}\sim a_0$ at large radius.
\subsection{2.4: A Brief Note on Source of Overdensity Angular Momentum}
During the overdensity growth, the overdense region can receive an angular momentum due to tidal fields from its neighbours. A common assumption is that angular momentum behaves as a small perturbation and does not affect the overdensity growth. In the small $\delta$ regime, the angular momentum growth based on tidal fields of
neighbours is modelled by Peebles \cite{peebles} in which the time dependent angular momentum per unit mass is given by the second order perturbation which grows as
\be
h(t) \propto C_P t^{5/3}.
\ee
where $C_P$ is some constant. White \cite{white1} obtains a first order perturbation from a flatten sphere (spheroid) in which the angular momentum grows as
\be
h(t) \propto C_W t,
\ee
where $C_W\gg C_P $ is also a constant. Casuso and Burkert in \cite{casuso}-\cite{burkert} propose that an expanding void can also provide a source of the protogalaxy's angular momentum. In
this scenario, the angular momentum transfer from the void causes an overdensity to breaks-away from its cosmological bacground and gravitationally collapses. 
In a simple model in which an overdensity monolithic collapses under Newtonian gravity, the angular momentum is assigned to mass shells at turnaround \cite{nusser}. 
\\\\
Although we work with a pure assumption following Mestel that the virialised sphere is given an uniform systematic angular velocity along some "z" direction such that it eqilibriates into a flat disk on its equatoral plane. A plausible scenario is that during evolution before the turnaround redshift, the overdensity is spherical and the physical radius is $r\propto a\varrho$ which is still expanding, the systematic angular velocity only grows slowly following Peebles' second order perturbation model so that the angular speed squared $\dot{\varphi}^2$ grows as
\be
\frac{h^2}{r^4} \propto C^2_P a(t).
\ee
After turnaround and during the relaxation process where the overdensity physical radius $r$ no longer expands, the spherical overdensity becomes spheroidal and the angular velocity starts to grow much faster following White's first order perturbation model. 
Here $\dot{\varphi}^2$ grows as
\be
\frac{h^2}{r^4}\propto C^2_W a(t)^3,
\ee
This scenario could produce an angular momentum growth model which mimicks the assumption in Mestel's analysis.
\subsection{3.1: Flat disk potential}
The Mestel disk is a well known model which has been explored to explain the flat rotational curve of rotating galaxies, some example can be found in Hoyle \cite{hoyle} and Davies \cite{davies}.
From Mestel \cite{mestel} Eq.(17)-(18), a spheroid of density $\rho_0$ (which is either uniform or follows a weak power law \cite{mestel} pp 564), with semi-axes $R_A$, $R_A$ and $(1-e^2)^{1/2} R_A$ exerts a gravitational acceleration at radius $r$
\begin{equation}
\ddot{r}=-\bigg (2\rho(1-e^2)^{1/2}R_A\bigg) \bigg(\frac{\pi G}{R_A}\bigg) \bigg(\frac{1}{e^3} (\sin^{-1}e-e(1-e^2)^{1/2})\bigg) r
\end{equation}
For a flat disk limit, $e\rightarrow 1$ and $\rho\rightarrow \infty$ such that $2\rho R_A(1-e^2)^{1/2}\rightarrow M_0=2\rho_0 R_A$ which is a surface mass density.
The disk radial acceleration at radius $r$ becomes
\be
\ddot{r}=- \bigg(\frac{3}{4} \pi \bigg) \frac{GM}{r^2}, \:\:\: M=\frac{4\pi}{3}\rho_0 r^3. 
\label{ddotr}
\ee
where $M$ is the enclosed (spheroid) mass at $r$. The disk acceleration differs from the acceleration of a sphere (or the point mass) by a factor of $3\pi/4$ and we could regard that the collapsing disk leads to a potential with an effective central mass $\frac{3\pi}{4} M$ at radius $r$ on the equatoral plane.
If the collapsing of a "simple" uniform sphere to disk occurs in VMOND, since the basis of our model is the competing free falling velocities Eq.(\ref{dotr1}), a consistent solution for the radial acceleration in Eq.(\ref{vmond}) is that its Newtonian gravitational mass is modified by a factor of $3\pi/4$ in both the Newtonian and non-Newtonian acceleration terms such that
\be
\ddot{r}=\frac{h^2}{r^3} - \frac{3\pi}{4}  \frac{GM}{r^2} -\sqrt{\frac{H(z)^2r}{2} }\sqrt{\frac{3\pi GM}{4r^2}}.
\label{dkvmond}
\ee
In terms of the virialised disk potential and the baryon density parameter $\Omega_b$, the rotation velocity takes the form
\be
V^4(R)=G\bigg(\frac{3 \pi}{4} M\bigg) \bigg( C_2(S)^2 H_0^2\Omega_b (1+z_{ta})^3R \bigg)\bigg(1+\frac{C_1(S)}{C_2(S)}\sqrt{\frac{3 \pi}{4}} n^{3/2}  \bigg)^2.
\label{realgamma}
\label{mestelgamma}
\ee
In spherical galaxy, we expect the central region is well described by a density profile $S=2$ (Singular isothermal sphere) to provide the fundamental plane in the Newtonian gravity dominant region. Away from the central region, the mass distribution could deviate from $S=2$, which could affects the rotational curve. In this work, we shall choose $S=2$ as a first approximation for the large radius considerations.
\section{3.2: The Milky Way Rotational Velocity}
Since we shall compare our modelled $a_0^{VM}$ and rotational velocity with observations, we need to confine the mass distribution of the Milky Way galaxy within the range of radius that we are considering.
\\\\
The Milky Way scale length is now  at $2.5kpc$ and the rotational curve out to $25kpc$ can be obtained to within small systematic errors \cite{eilers}, \cite{ou}.
Earlier work estimates that the stellar mass inside galaxy centre to solar radius $R_0=8.5kpc$ is given by $M_*(R_0)=6.07\times 10^{10}M$ \cite{licquia}-\cite{licquia2}. From Gaia DR3 measurements \cite{ou}, the Milky Way mass budget is given by bulge + exponential disk+ gas+ dust equals $M=6.2\times 10^{10}M_{\odot}$. McGaugh \cite{mcgaugh2015} uses a Milky Way mass $7\times 10^{10} M_{\odot}$ and Nicastro et al. \cite{nicastro} gives an upper limit for Milky Way mass at $ 10\times 10^{10} M_{\odot}$. Our Milky Way mass estimate comes from the mass of bulge and exponential disk \cite{kalberla} plus the gas and dust in \cite{kalberla2}, which we call $M_R$ where $R$ is the radius. At $R=16kpc$, our mass estimate is
\begin{equation}
M_{16kpc} =M_{16}= 7.4 \times 10^{10}M_{\odot}.
\end{equation}
which is within the range of \cite{nicastro}. Detailed modelling Milky Way rotational curves have been done by many, see Sofue \cite{sofue} for CDM modelling, McGaugh \cite{mcgaugh2015} for using the empirical Radial Discrepancy Acceleration Relation (RDAR), Moffat \cite{moffat4} for using MOG. In these modelling the disk is modelled by variations of an exponential disk. 
Upto galactic centre-solar radius $R_0=8kpc$, Newtonian gravity based on the exponential disk model, for example Freeman disk \cite{freeman}, works well. 
\\\\
For $8kpc-16kpc$, the observed rotational curve is approximately flat, which different modelling can match observations with some form of non-Newtonian ingredients. Specifically, dark matter model uses massive halo by Sofue \cite{sofue}, McGaugh \cite{mcgaugh2015} uses empircal RDAR and Moffat \cite{moffat4} uses MOG. All these models could match observation given some parameter fitting. For canonical MOND, this radius range lies in an interpolation region where there is a significant parameter flexibility to match data. 
\\\\
Beyond $R=16kpc$ which is far from the galactic mass centre, the central gravity can be represented by a central point mass. In canonical MOND, this radius is approaching the deep MOND region which is found to be problematic by \cite{chan}, \cite{coquery}. In \cite{moffat} the MOG model predicts that there should be a minimum of $30\%$ missing baryons and in CDM fit, the descend of the rotational vurve is not as steep as the data observed.
\\\\
Since in our virialised potential Eq.(\ref{psi}), the short distance potential is Newtonian which should support similar physics of the Freeman disk. In the flat rotational curve region $R_0-2R_0$, Davies  \cite{davies} finds that the Newtonian Mestel disk works reasonably well.  It is the decreasing rotational curve range from radius $R>2R_0=16kpc$ that our model prediction can be truly tested. Here, our aim is to compare Rotational curve data from \cite{eilers}-\cite{ou} with our model prediction.
\\\\
Given the exponential profile of the gas, we assume that $M_{16kpc}$ is a good mass estimate inside radius $17.77kpc$.
In our Milky way, for easy comparison $a_0^{VM}$ with $a_0$, we write the phenomenological MOND acceleration $a_0$ in multiples of $\frac{1}{2}H_0^2 r_{0}$, where $r_0=2.5kpc$ is the Milky Way scale length given by \cite{ou},
\be
a_0=\frac{1}{6}cH_0=\bigg(\frac{1}{2}H_0^2r_{0}\bigg)\gamma;\:\:\gamma=\frac{c/H_0}{3\times 2.5kpc}=5.33\times 10^5,
\label{a0}
\ee
At $S=2$, using Eq.(\ref{vsigma11}), we obtain
\be
\gamma^{VM}=\frac{3\pi}{16} \frac{R}{r_0}\Omega_b (1+z_{ta})^3( \bigg(1+2\sqrt{\frac{3 \pi}{4}} n^{3/2}  \bigg)^2.
\label{realgamma1}
\ee
so that the modelled value of $\gamma^{VM}$ in Eq.(\ref{realgamma1}) can be compared easily with the canonical MOND $\gamma$ value in Eq.(\ref{a0}). 
\\\\
Similar to the case in \cite{wong2}, we take $\delta=2.8\times 10^{-3}$ ($A_0=60$) which leads to $z_{ta}=29$, $r_{ta}=44.43 kpc$,  $n=\frac{44.43}{r}$ and using Eq.(\ref{a0})-Eq.(\ref{realgamma1}) we obtain $a_0^{VM}(R)$ as follows
\be
a_0^{VM}(17.7kpc)=1.79a_0, \:\:\: a_0^{VM}(19.71kpc)=1.46a_0,\:\: a_0^{VM}(25.92kpc)=0.983a_0\:\: a_0^{VM}(27.3kpc) =0.856a_0, 
\ee
This model prediction Eq.(\ref{vsigma11}) is an BTFR with $a_0^{VM}(R) \sim O(a_0)$ for $R >2R_0$.  However our model $a_0^{VM}(R)$ is not a constant but decreases as radius $R$ increases. So far, they are pure theoretical predictions, but the non-canonical $a_0$ feature is confirmed rexperimentally from Gaia DR3 data by \cite{chan}, \cite{coquery}.
\\\\
Finally, we calculate our model rotational velocity $V(R)$ from Eq.(\ref{vsigma11}) using $z_{ta}=29$, $z_{ta}=24$ and $z_{ta}=34$ and compare with the $V_c $ data given in \cite{eilers} and \cite{ou}. The results are shown in Fig.1. 
\\\\
We see that our model rotational velocity $V(R)$ for $z_{ta}=29$ matches the Gaia DR3 results very closely.  It is also worth noting that model prediction continues to match data closely for the range of turnaround redshifts  $ 24\leq z_{ta}\leq 33$. This means that our prediction remains viable as long as the choice of initial overdensity for the Milky Way formation lies in $1.94\times 10^{-3} <\delta_{int} < 3.5 \times 10^{-3}$ whch leads to the desirable turnaround redshift. This is a surprising result,  given the simplie Mestel disk formation assumption made in our modelling. We wish to point out that, for better model prediction, immediately after recombination we need to take into account of the effects of radiation and other non-baryonic background components should they exist. 
\\\\ 
The above result suggests that the non-Newtonian potential from our new metric could provide a new baseline from which to model galaxy formation from an early time overdensity. It could also explain how the Tully-Fisher relation arises as well as providing a first principle origin for the phenomenological MOND acceleration.
\section{Summary and Discussion}
We consider a simple model of rotating galaxy formation, in which a Milky Way mass spherical overdensity under VMOND evolves to virialisation similar to the work of \cite{wong2}. Following Mestel \cite{mestel}, in the virialised sphere we assume it is given an uniform systematic angular velocity (where angular momentum is conserved at particle level) such that the sphere equilibriates to a thin disk. In Mestel \cite{mestel}, working in Newtonian gravity, the spheroid gravitational acceleration on the equatoral plane will change to include a factor $3\pi/4$ multiplied to the graviational mass in the thin disk limit. In VMOND acceleration, which combines a Newtonian and MOND-like acceleration, a consistent solution is to include the same factor $3\pi/4$ to the gravitational mass in both the Newtonian and the non-Newtonian acceleration in the disk acceleration. 
\\\\
From the virialised disk potential, using $S=2$, at $z_{ta}=29$ we find that at radius $17.7-27.3kpc$, the predicted MOND acceleration is $a_0^{VM}(R)\sim O (a_0)$ but its value reduces as radius increases. We also calculate the rotating velocity in the same radius range and find that they match results from Gaia DR3 to within error bars. This result remains valid for a range $24\leq z_{ta} \leq 33$.
\\\\
The result in this work and \cite{wong2} support the suggestion that the VMOND acceleration from the new metric in \cite{wong}, could explain the mass discrepancy problem at galactic scales. The missing mass problems at very large scales, such as in CMB angular power spectrum and Matter Power Spectrum require further input which  are addressed elsewhere \cite{wong3} and \cite{wong4}.
\subsection{Appendix 1: The late time solar system particle orbit in the VMOND potential}
It will be instructive to see the impact of the non-Newtonian acceleration in the solar system scale.  First we consider a VMOND orbit formed at late time.
It is useful to study the orbit of a slow speed particle of unit mass around a fixed central mass M
with specific angular momentum $h$ under the influence of a central potential with polar coordinate $(r,
\varphi)$ (e.g. see \cite{daboul}]) that satisfies
\begin{equation}
\varphi= \int_{r_0}^{r} \frac {r^{-1} dr} {\sqrt{ \frac{2}{h^2}(E-V)r^2-1}}
\label{phi0}
\end{equation}
where E is the total energy. Eq.(\ref{phi0}) becomes
\begin{equation}
\varphi =\int_{r_{0}}^{r} \frac {r^{-1} dr} {\sqrt { \frac{2}{h^2}(GM/r+E-H(z)\sqrt{2GMr} + H(z)^2r^2/2)  r^2
-1}} =\int_{r_{0}}^{r} \frac {r^{-1} dr} {\sqrt { \frac{2}{h^2}(\frac{GM}{r}
(1-\sqrt{\frac{\rho_H(z)}{\rho_b(r)}} )^2 +E)-1}}.
\label{phi1}
\end{equation}
where $\rho_H(z)$, $\rho_b(r)$ are given in Eq.(\ref{potl44}) and we define $\alpha$ as in Eq.(\ref{potl44}).
We stress that we are no longer considering a protogalaxy growth, but only a point particle in a VMOND potential starting from matter dominant epoch to the present time.
\\\\
We follow the
usual parametrisations by taking a length scale $l$  with $\frac{h^2}{2GM} =l $, $E_0 = \frac{El}{GM}$ and
eccentricity $\epsilon^2=1+4E_0$.
\\\\
Rearranging Eq. (\ref{phi1}) we obtain
\begin{equation}
\varphi =\int_{r_{0}}^{r} \frac {2lr^{-2} dr} {\epsilon \sqrt { 1+ \frac{1}{\epsilon^2}(\alpha^2-1)
-\frac{1}{\epsilon^2}\bigg(\alpha-\frac{2l}{r}\bigg)^2   }} =\int_{r_{0}}^{r} \frac {2lr^{-2} dr} {\epsilon \sqrt
{ 1- \frac{1}{\epsilon^2} -\frac{1}{\epsilon^2}\bigg(\frac{2l}{r}\bigg)^2 +\bigg(\frac{2l}{\epsilon
r}\bigg)\bigg(\frac{2\alpha}{\epsilon}\bigg)   }}
\label{phi2}
\end{equation}
Setting
\begin{equation}
\epsilon x =- \frac{2l} {r}
\end{equation}
we obtain
\begin{equation}
\varphi =\int  \frac{dx}{ \sqrt{ \bigg(1-\frac{1}{\epsilon^2} \bigg)-\frac{2\alpha}{\epsilon}x-x^2 }}
\label{phi3}
\end{equation}
Although $\alpha (r)$ depends on $r$ and $z$, its variation is based on a cosmological time scale.  We obtain
\begin{equation}
\bigg(1+\frac{\alpha^2-1}{\epsilon^2}\bigg)^{1/2}\epsilon\: sin\varphi \approx
\alpha-\frac{2l}{r},
\label{orbit}
\end{equation}
That is, in Keplerian form
\begin{equation}
 r=\frac{2l'}{1-\epsilon' sin\varphi}
\label{r}
\end{equation}
where
\be
l' = \frac{l}{\alpha},\,\,\,\,\,\,\,\,\,\epsilon' =
\frac{\epsilon}{\alpha}\bigg(1+\frac{\alpha^2-1}{\epsilon^2}\bigg)^{1/2}.\:\:\:\: \alpha =\bigg(1-\sqrt{\frac{\rho_H(z)}{\rho_b(r)}}\bigg)^2.
\label{rr}
\ee
\\\\
Firstly, we consider a particle orbit which is formed only recently in the cosmological constant dominant epoch, where $H(z) =H_0=73kms^{-1} Mpc^{-1}$ is nearly constant and ($\alpha \sim 1$).. Specifically, we consider the Earth's orbit around the Sun with radius at $1 AU=1.496\times
10^{11} m$, we have $\rho_c=3H_0^2/8\pi G$ and
\be
\frac{\rho_H}{\rho_b(r)}=\frac{\rho_c}{\rho_b(r)}=\frac{1.06 \times 10^{-26} kgm^{-3}}{M_{\odot}/\frac{4\pi }{3} (AU)^3} =\frac{1.06\times
10^{-26}}{1.42\times 10^{-4}}\sim 7.46\times 10^{-23},
\ee
\be
\sqrt{\frac{\rho_H} {\rho_b(r) }} =8.64 \times 10^{-12},\:\:\:\:2\sqrt{\frac{\rho_H} {\rho_b(r) }}  < 1.728\times 10^{-11}.
\ee
\be
l'=l \bigg(1+2\sqrt{\frac{\rho_H} {\rho_b(r) }} +O(10^{-22})\bigg)
\ee
The VMOND correction to the closest approach is a ($+1.728\times 10^{-11})$ effect. In comparison, the Viking observation
uncertainties for Earth and Mars orbits are $100m$ and $150m$ respectively which are accurate to $O(10^{-10})$ \cite{shapiro}-\cite{anderson3}. More specifically, the uncertainty for Earth's orbit is $66.84\times 10^{-11}$ which is just under two orders of magnitude larger than the VMOND correction to the closest approach. Therefore, if a planetary orbit is formed recently where $H(z)=H_0$, the VMOND correction is still well within our current observational limits. It may be possible to see directly whether this VMOND effect exists if the orbital measurement uncertainty is improved.
\subsection{Appendix 2: VMOND at Wide-Binary scale}
In the MOND scheme, when the Newtonian acceleration magnitude $g_N$ is $g_N\ll a_0$, where $a_0=1.2\times 10^{-10} ms^{-2}$ is a canonical scale, the MOND gravitational acceleration takes the value
\be
\ddot{r}_{M} =-\sqrt{g_N a_0}.
\label{rM}
\ee
The Wide Binaries parameters are $r=7-30\:KAU$ with typical mass at $1-2 M_{\odot}$. Take $r=0.1pc$, $M=2M_{\odot}$, the Newtonian acceleration magnitude is
\be
\frac{GM}{r^2}=\frac{6.67\times 10^{-11} \times 4\times 10^{30}}{(3\times 10^{15})^2}ms^{-2}=2.96 \times 10^{-11} ms^{-2},
\ee
which is smaller than the MOND acceleration $a_0$, where one would expect to see effects of Eq.(\ref{rM}).
More specifically, comparing their accelerations, without taking into accounts the other MOND effects such as external field effect, 
\be
\gamma_{grav}= \frac{\ddot{r}_{M}}{g_N}=\sqrt{\frac{1.2\times 10^{-10}}{2.96\times 10^{-11}}} =2.01.
\label{eq1}
\ee
One would therefore expect to see effect of MOND at Wide Binary scale. In Banik et al. \cite{banik}, for the observed acceleration $\ddot{r}_{obs}$, they find that the data supports 
\be
\gamma_{grav} =\frac{\ddot{r}_{obs}}{g_N} \simeq 1.
\ee
to high precision. Since Eq.(\ref{eq1}) indicates the theoretical MOND correction to Newtonian $\alpha_{grav}$ is $O(1)$, the Banik et al. analysis if further confirmed will be seriously problematic for the canonical MOND paradigm. 
\\\\
In VMOND, the gravitational acceleration
\be
\ddot{r}_{VM}=-\frac{GM}{r^2}-H\sqrt{\frac{GM}{2r}} +\frac{\ddot{a}}{a}r.
\ee
\be
\gamma_{grav}=\frac{\ddot{r}_{VM}}{g_N} =1+\sqrt{\frac{H^2r^3}{2GM}}-\frac{\ddot{a}}{a}\frac{r^3}{GM}.
\ee
We take
\be
H=H_0=2.43\times 10^{-18} s^{-1},\:\:\: M=2M_{\odot},\:\:\: r=0.1pc.
\ee
\be
\frac{GM}{r^3}=\frac{6.67\times 10^{-11} \times 4\times 10^{30}}{ (3\times 10^{15})^3}\simeq 10^{-26} s^{-2}.
\ee
\be
\frac{H^2r^3}{2GM}=2.95\times 10^{-10}.
\ee
From a cosmology without dark matter particle \cite{wong3},
\be
\frac{\ddot{a}}{a}=H_0^2\bigg (\Omega_{\Lambda}-\frac{3}{4} (2\sqrt{\Omega_b\Omega_{\Lambda}})-\frac{1}{2}\Omega_b\bigg).
\ee
\be
\frac{\ddot{a}}{a}\frac{r^3}{GM}=\frac{(2.43\times 10^{-18})^2 \times 0.38}{10^{-26}}=2.24\times 10^{-10}.
\ee
so that
\be
\gamma_{grav} =\frac{\ddot{r}_{VM}}{g_N}=1+1.71\times 10^{-5}-2.24\times 10^{-10}.
\ee
which supports a Newtonian acceleration dominance scenario in a Wide Binary scale system..

\end{document}